\documentclass[twocolumn,tighten,times]{aastex61}
\usepackage{amssymb}
\usepackage{amsmath}
\usepackage{amsfonts}
\usepackage{float}
\usepackage{relsize}
\usepackage{verbatim}
\usepackage{color}
\usepackage{comment}
\usepackage{graphicx}
\usepackage{bm}
\usepackage{enumitem}

\usepackage{silence}
\newcommand{\todoblank}[1]{}

\newcommand{\addedA}[1]{{#1}}
\newcommand{\addedB}[1]{{#1}}

\usepackage{xspace}
\graphicspath{{figures/}}

\bibpunct[; ]{(}{)}{;}{a}{}{,} 

\def\be{\begin{equation}}
\def\ee{\end{equation}}
\def\ba{\begin{eqnarray}}
\def\ea{\end{eqnarray}}

\newcommand{\SN}{\ensuremath{\mathrm{S/N}}} 
\newcommand{\DM}{\ensuremath{\mathrm{DM}}} 
\newcommand{\DDM}{\ensuremath{\mathrm{\Delta DM}}} 
\newcommand{\Dne}{\Delta n_e} 
\newcommand{\Halpha}{\ensuremath{\mathrm{H\alpha}}}

\newcommand{\Rearth}{\ensuremath{\mathrm{r_\oplus}}}
\newcommand{\Dnud}{\Delta\nu_{\mathrm{d}}}
\newcommand{\taud}{\tau_{\mathrm{d}}}
\newcommand{\Dttwo}{\Delta t_{2,1400}}
\newcommand{\Dtfour}{\Delta t_{4,1400}}

\newcommand{\citeBrook}{P. Brook. et al. \addedB{(submitted)}\xspace}
\newcommand{\citeMadison}{(\addedA{a global fit over all NANOGrav DM measurements; }D.~R.~Madison et al. in prep.)\xspace}

\def\app#1#2{%
  \mathrel{%
    \setbox0=\hbox{$#1\sim$}%
    \setbox2=\hbox{%
      \rlap{\hbox{$#1\propto$}}%
      \lower1.1\ht0\box0%
    }%
    \raise0.25\ht2\box2%
  }%
}

\shorttitle{Interstellar Medium Events in PSR~J1713+0747}
\shortauthors{Lam et al.}

\begin{document}

\title{
A Second Chromatic Timing Event of Interstellar Origin toward PSR J1713+0747
}

\author[0000-0003-0721-651X]{M.\,T.\,Lam}
\affiliation{Department of Physics and Astronomy, West Virginia University, P.O. Box 6315, Morgantown, WV 26506, USA}
\affiliation{Center for Gravitational Waves and Cosmology, West Virginia University, Chestnut Ridge Research Building, Morgantown, WV 26505, USA}
\author{J.\,A.\,Ellis}
\affiliation{Department of Physics and Astronomy, West Virginia University, P.O. Box 6315, Morgantown, WV 26506, USA}
\affiliation{Center for Gravitational Waves and Cosmology, West Virginia University, Chestnut Ridge Research Building, Morgantown, WV 26505, USA}
\author{G.\,Grillo}
\affiliation{Department of Astronomy, Cornell University, Ithaca, NY 14853, USA}
\author{M.\,L.\,Jones}
\affiliation{Department of Physics and Astronomy, West Virginia University, P.O. Box 6315, Morgantown, WV 26506, USA}
\affiliation{Center for Gravitational Waves and Cosmology, West Virginia University, Chestnut Ridge Research Building, Morgantown, WV 26505, USA}
\author{J.\,S.\,Hazboun}
\affiliation{Center for Advanced Radio Astronomy, University of Texas Rio Grande Valley, 1 West University Dr., Brownsville, TX 78520}
\affiliation{Division of Physical Sciences, School of STEM, University of Washington Bothell, Box 358500, 18115 Campus Way NE, Bothell, WA 98011-8246}
\author{P.\,R.\,Brook}
\affiliation{Department of Physics and Astronomy, West Virginia University, P.O. Box 6315, Morgantown, WV 26506, USA}
\affiliation{Center for Gravitational Waves and Cosmology, West Virginia University, Chestnut Ridge Research Building, Morgantown, WV 26505, USA}
\author{J.\,E.\,Turner}
\affiliation{Center for Gravitation, Cosmology and Astrophysics, Department of Physics, University of Wisconsin-Milwaukee, P.O. Box 413, Milwaukee, WI 53201, USA}
\author{S.\,Chatterjee}
\affiliation{Department of Astronomy, Cornell University, Ithaca, NY 14853, USA}
\author{J.\,M.\,Cordes}
\affiliation{Department of Astronomy, Cornell University, Ithaca, NY 14853, USA}
\author{T.\,J.\,W.\,Lazio}
\affiliation{Jet Propulsion Laboratory, California Institute of Technology, 4800 Oak Grove Drive, Pasadena, CA 91109, USA}
\author{M.\,E.\,DeCesar}
\affiliation{Department of Physics, Lafayette College, Easton, PA 18042, USA}
\author{Z.\,Arzoumanian}
\affiliation{Center for Research and Exploration in Space Science and Technology and X-Ray Astrophysics Laboratory, NASA Goddard Space Flight Center, Code 662, Greenbelt, MD 20771, USA}
\author{H.\,Blumer}
\affiliation{Department of Physics and Astronomy, West Virginia University, P.O. Box 6315, Morgantown, WV 26506, USA}
\affiliation{Center for Gravitational Waves and Cosmology, West Virginia University, Chestnut Ridge Research Building, Morgantown, WV 26505, USA}
\author{H.\,T.\,Cromartie}
\affiliation{University of Virginia, Department of Astronomy, P.O. Box 400325, Charlottesville, VA 22904, USA}
\author{P.\,B.\,Demorest}
\affiliation{National Radio Astronomy Observatory, 1003 Lopezville Rd., Socorro, NM 87801, USA}
\author[0000-0001-8885-6388]{T.\,Dolch}
\affiliation{Department of Physics, Hillsdale College, 33 E. College Street, Hillsdale, Michigan 49242, USA}
\author{R.\,D.\,Ferdman}
\affiliation{School of Chemistry, University of East Anglia, Norwich, NR4 7TJ, United Kingdom}
\author{E.\,C.\,Ferrara}
\affiliation{NASA Goddard Space Flight Center, Greenbelt, MD 20771, USA}
\author[0000-0001-8384-5049]{E.\,Fonseca}
\affiliation{Department of Physics, McGill University, 3600  University St., Montreal, QC H3A 2T8, Canada}
\author{N.\,Garver-Daniels}
\affiliation{Department of Physics and Astronomy, West Virginia University, P.O. Box 6315, Morgantown, WV 26506, USA}
\affiliation{Center for Gravitational Waves and Cosmology, West Virginia University, Chestnut Ridge Research Building, Morgantown, WV 26505, USA}
\author{P.\,A.\,Gentile}
\affiliation{Department of Physics and Astronomy, West Virginia University, P.O. Box 6315, Morgantown, WV 26506, USA}
\affiliation{Center for Gravitational Waves and Cosmology, West Virginia University, Chestnut Ridge Research Building, Morgantown, WV 26505, USA}
\author{V.\,Gupta}
\affiliation{Centre for Astrophysics and Supercomputing, Swinburne University of Technology, P.O. Box 218, Hawthorn, Victoria 3122, Australia}
\author{D.\,R.\,Lorimer}
\affiliation{Department of Physics and Astronomy, West Virginia University, P.O. Box 6315, Morgantown, WV 26506, USA}
\affiliation{Center for Gravitational Waves and Cosmology, West Virginia University, Chestnut Ridge Research Building, Morgantown, WV 26505, USA}
\author{R.\,S.\,Lynch}
\affiliation{Green Bank Observatory, P.O. Box 2, Green Bank, WV 24944, USA}
\author{D.\,R.\,Madison}
\affiliation{National Radio Astronomy Observatory, 520 Edgemont Road, Charlottesville, VA 22903, USA}
\author[0000-0001-7697-7422]{M.\,A.\,McLaughlin}
\affiliation{Department of Physics and Astronomy, West Virginia University, P.O. Box 6315, Morgantown, WV 26506, USA}
\affiliation{Center for Gravitational Waves and Cosmology, West Virginia University, Chestnut Ridge Research Building, Morgantown, WV 26505, USA}
\author[0000-0002-3616-5160]{C.\,Ng}
\affiliation{Department of Physics and Astronomy, University of British Columbia, 6224 Agricultural Road, Vancouver, BC V6T 1Z1, Canada}
\author[0000-0002-6709-2566]{D.\,J.\,Nice}
\affiliation{Department of Physics, Lafayette College, Easton, PA 18042, USA}
\author[0000-0001-5465-2889]{T.\,T.\,Pennucci}
\affiliation{Institute of Physics, E\"{o}tv\"{o}s Lor\'{a}nd University, P\'{a}zm\'{a}ny P. s. 1/A, Budapest 1117, Hungary}
\affiliation{Hungarian Academy of Sciences MTA-ELTE Extragalactic Astrophysics Research Group, 1117 Budapest, Hungary}
\author[0000-0001-5799-9714]{S.\,M.\,Ransom}
\affiliation{National Radio Astronomy Observatory, 520 Edgemont Road, Charlottesville, VA 22903, USA}
\author[0000-0002-6730-3298]{R.\,Spiewak}
\affiliation{Centre for Astrophysics and Supercomputing, Swinburne University of Technology, P.O. Box 218, Hawthorn, Victoria 3122, Australia}
\author[0000-0001-9784-8670]{I.\,H.\,Stairs}
\affiliation{Department of Physics and Astronomy, University of British Columbia, 6224 Agricultural Road, Vancouver, BC V6T 1Z1, Canada}
\author{D.\,R.\,Stinebring}
\affiliation{Department of Physics and Astronomy, Oberlin College, Oberlin, OH 44074, USA}
\author{K.\,Stovall}
\affiliation{National Radio Astronomy Observatory, 1003 Lopezville Rd., Socorro, NM 87801, USA}
\author{J.\,K.\,Swiggum}
\affiliation{Center for Gravitation, Cosmology and Astrophysics, Department of Physics, University of Wisconsin-Milwaukee, P.O. Box 413, Milwaukee, WI 53201, USA}
\author{S.\,J.\,Vigeland}
\affiliation{Center for Gravitation, Cosmology and Astrophysics, Department of Physics, University of Wisconsin-Milwaukee, P.O. Box 413, Milwaukee, WI 53201, USA}
\author{W.\,W.\,Zhu}
\affiliation{National Astronomical Observatories, Chinese Academy of Science, 20A Datun Road, Chaoyang District, Beijing 100012, China}
\affiliation{Max Planck Institute for Radio Astronomy, Auf dem H\"{u}gel 69, D-53121 Bonn, Germany}

\correspondingauthor{M.\,T.\,Lam}
\email{michael.lam@mail.wvu.edu}

\begin{abstract}

The frequency dependence of radio pulse arrival times provides a probe of structures in the intervening media. \citet{Demorest+2013} \addedA{was the first to show} a short-term ($\sim$100-200~days) reduction in the electron content along the line of sight to PSR~J1713+0747 in data from 2008 (approximately MJD~54750) based on an apparent dip in the dispersion measure of the pulsar. We report on a similar event in 2016 (approximately MJD~57510), with average residual pulse-arrival times $\approx\!-3.0,-1.3$, and $-0.7~\mu\mathrm{s}$ at 820, 1400, and 2300~MHz, respectively. Timing analyses indicate \addedA{possible} departures from the standard $\nu^{-2}$~dispersive-delay dependence. We discuss and rule out a wide variety of \addedB{potential} interpretations. We find the likeliest scenario to be lensing of the radio emission by some structure in the interstellar medium, which causes multiple frequency-dependent pulse arrival-time delays.

\end{abstract}


\section{Introduction}

Precise timing of recycled millisecond pulsars provides access to a number of stringent tests of fundamental physics \citep[e.g.,][]{wnrs2007,Will2014,Kramer2016}.  A standard component of precision timing models is a frequency-dependent dispersive delay proportional to the dispersion measure ($\DM=\int n_e\,\,ds$), the integral of the electron density $n_e$ along the line of sight \citep[LOS;][]{handbook}. Temporal variations in the measured dispersive delay have been observed and interpreted as being caused by: LOS changes in $n_e$, Earth-pulsar distance and direction changes, solar wind fluctuations, ionospheric electron content variations, contamination from refraction, and more \citep[e.g.,][]{fc90,Ramachandran+2006,Keith+2013,LamDMt,NG9DM}. 
\addedA{High-precision observations of pulse time-of-arrival (TOA) timeseries over a wide range of frequencies allow for the measurement of a number of propagation effects.  

\addedB{In this paper,} we refer to frequency-independent phenomena (such as pulse spin, pulsar-Earth distance variation, etc) as achromatic.  We refer to frequency-dependent phenomena (including $\propto\nu^{-2}$ interstellar dispersion where $\nu$ is the radio frequency, along with phenomena which depend on $\nu$ in other ways) as chromatic.} Pulsar timing arrays (PTAs) allow for high-sensitivity observations of many types of \addedA{chromatic} TOA variations over many LOSs \citep{Stinebring2013}.

\begin{figure*}[t!]
\centering
\includegraphics[width=0.75\textwidth]{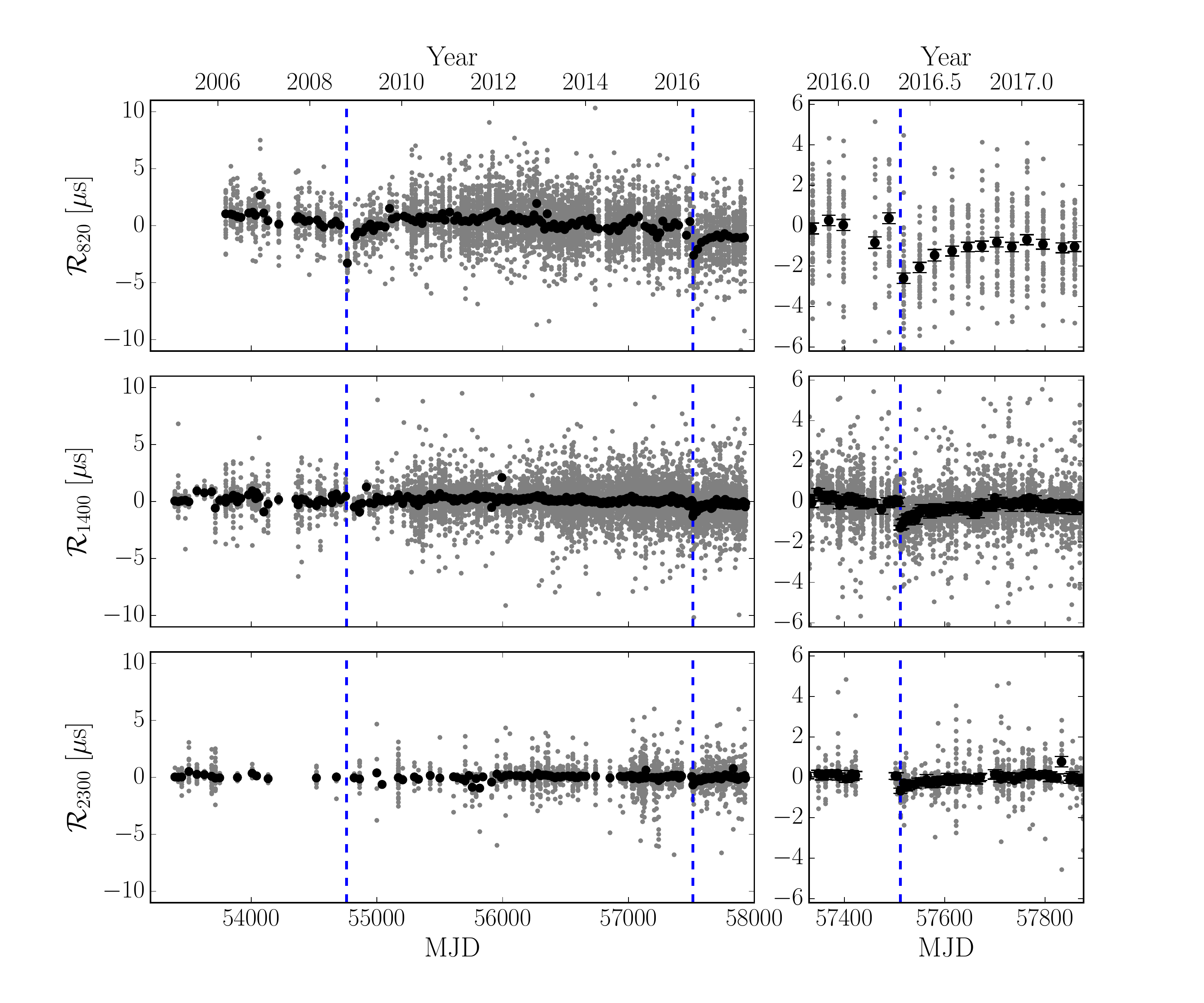}
  \caption{\footnotesize Left: Timing residuals $\mathcal{R}$ (TOAs minus simplified timing model; see text) as a function of frequency for the 820 (top), 1400 (middle), and 2300 (bottom) MHz bands. The gray dots show the per-frequency-channel residuals with the weighted mean \addedA{for each frequency channel} subtracted while the black dots show the epoch-averaged residuals \citep{NG9WN}. The dashed lines indicate the start times of both events (measured from the higher-cadence 1400~MHz data). Right: A zoom-in of the residuals for the second event with errors \addedA{shown on the individual and the averaged residuals}. Note the different y-axis scale.}
\label{fig:wnresids}
\end{figure*}

\setcounter{footnote}{0}

PSR~J1713+0747 is one of the best-timed pulsars observed by the North American Nanohertz Observatory for Gravitational Waves (NANOGrav; \citealt{McLaughlin2013}). Previously, a \addedA{chromatic timing event -- that is, a relatively sudden, frequency-dependent change in timing properties -- }was seen in the TOAs starting at approximately MJD~54750, interpreted as a DM drop of $\approx\!6\times 10^{-4}$~pc~cm$^{-3}$  and lasting for $\sim$100-200 days before returning to the previous DM value; the event was seen in other datasets as well \citep{Demorest+2013,Keith+2013,Desvignes+2016,NG9DM}.

We report on a second chromatic-timing event occurring 7.6 years after the first event. We discuss our radio observations of PSR~J1713+0747 in \S\ref{sec:observations}, \addedA{timing analyses in \S\ref{sec:analysis}, and pulse-profile analyses in \S\ref{sec:otherradio}. The line of sight in infrared and optical wavelengths is described in \S\ref{sec:multiwavelength}}. Possible interpretations of the two events are given in \S\ref{sec:interpretations}, and we briefly discuss the results and implications for future timing observations in \S\ref{sec:discussion}.

\begin{figure*}[t!]
\centering
\includegraphics[width=0.90\textwidth]{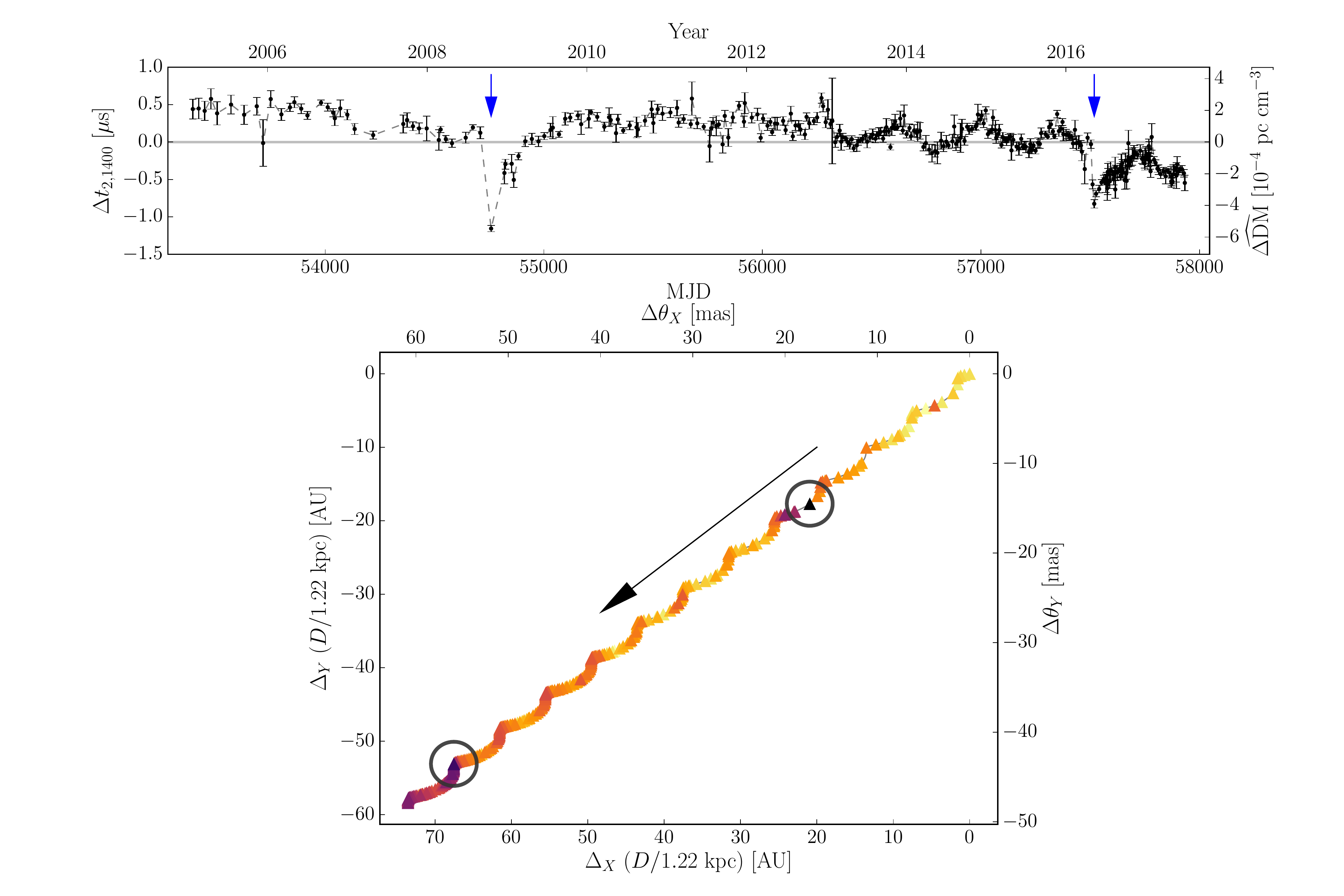}
  \caption{\footnotesize Top: Modeled timeseries of $\Dttwo\propto\!\nu^{-2}$ delays using \textsc{Tempo}, scaled to 1400~MHz. Assuming the delays are purely dispersive, the equivalent estimated $\Delta\widehat{\DM}$ is shown on the right y-axis. Errors are \addedA{1$\sigma$ and} given from the generalized least-squares fit \addedA{that reflects the uncertainty in the relative epoch-to-epoch values and does not include the systematic uncertainty in the absolute DM}. Bottom: Apparent trajectory of the pulsar on the sky (direction given by arrow) with the $\Dttwo$ values color coded, with darker values indicating smaller $\Dttwo$. The two events have been circled for clarity (arrows in the top panel). The spatial scales have been set for the pulsar's distance of 1.22~kpc though the curve's shape will depend slightly on the distance $D$ considered.}
\label{fig:DMX}
\end{figure*}

\section{Observations}
\label{sec:observations}

\addedA{Here we describe our observations of PSR~J1713+0747. These data are part of the preliminary NANOGrav 12.5-year data release. This data release} will include new methodologies for pulsar timing and comparisons between the procedures; however, for the present work, we use procedures previously used in the NANOGrav 11-year Data Set \citep[NG11;][]{NG11yr}, which are discussed below along with some modifications. A more detailed account of the methods here can be found in NG11.


PSR J1713+0747 was observed using the Arecibo Observatory (AO) and the Green Bank Telescope (GBT). We observed pulse profiles with AO \addedA{at 1400 and 2300~MHz} using the \addedB{Arecibo Signal Processor} backend \addedA{(\addedB{ASP;} up to 64~MHz bandwidth)} and then the \addedA{larger-bandwidth} \addedB{Puerto Rico Ultimate Pulsar Processing Instrument} backend \addedA{(\addedB{PUPPI;} up to 800~MHz bandwidth)} \addedB{since} 2012 \citep{NG9yr}. At GBT, we used the nearly identical \addedB{Green Bank Astronomical Signal Processor (GASP)} backend and then the \addedB{Green Bank Ultimate Pulsar Processing Instrument (GUPPI)} backend after 2010 \addedA{to observe at 820 and 1400~MHz. We observed at these multiple frequencies in order to determine the DM value per epoch. The ASP and GASP profiles were observed with 4 MHz frequency channels while the GUPPI and PUPPI profiles were observed with channels ranging from 1.5 to 12.5~MHz depending on the receiver system used. Originally we observed at an approximately monthly cadence at both telescopes but switched to weekly starting in 2013 at the GBT and 2015 at AO.}

Profiles were flux and polarization calibrated and radio-frequency interference was removed. We generated TOAs with a template-matching procedure \citep{Taylor1992} \addedB{via \textsc{psrchive}\footnote{\url{http://psrchive.sourceforge.net}} \citep{Hotan+2004,vSdo2012}}. Outlier TOAs were removed via an automated method where we removed TOAs with the probability of being \addedA{within the uniform outlier distribution} $\ge10\%$ \citep{Outlier,NG11yr}.

\section{Timing Analyses}
\label{sec:analysis}

\addedA{Here we describe analyses performed on the TOAs to investigate these events. We first demonstrate the presence of the events using a fixed timing model without any \addedB{time-varying} chromatic terms. Next we fit a timing and noise model following standard pulsar timing practices. Finally, we tried to introduce physical and phenomenological model components for the chromatic timing variations.}

\subsection{Fixed Achromatic Timing Model}
\label{sec:fixedmodel}

Figure~\ref{fig:wnresids} shows the timing residuals, TOAs minus a fixed and simplified timing model modified from NG11 (ending prior to 2016), from measurements taken in the 820, 1400, and 2300~MHz frequency bands.  Initially we used the NG11 parameters to avoid contamination of timing- and noise-model parameters by the \addedA{second} event. The first event was previously modeled as time-varying DM fit \addedA{per epoch using TOAs from the} three narrow frequency bands \addedA{\citep{Demorest+2013}}. Our simplified \addedA{fixed model} only included spin, astrometric, binary, and telescope parameter terms along with a constant DM, i.e., we removed all parameters describing the \addedA{step-wise} variations in DM (DMX), frequency-dependent pulse-profile-evolution delays (FD), modifications to the TOA errors (EFAC, EQUAD, ECORR), and any excess red noise (RNAMP, RNIDX). We included noise-model parameters measured in \citet{NG9WN} to account for pulse jitter and scintillation noise \addedA{on the TOA uncertainties for proper weighting later when averaging within epochs}. We excised \addedA{remaining residuals from profiles with low signal-to-noise ratio (S/N)} with errors $\ge3~\mathrm{\mu s}$\footnote{Pulse $\mathrm{S/N}\approx4$.}. We then calculated timing residuals, again holding the parameters fixed. For each \addedA{narrow} frequency channel \addedA{(not each frequency band)} we subtracted the weighted-mean average from the TOAs \addedA{to account for unknown delays from frequency-dependent pulse-profile evolutions \citep{NG9WN}} and then computed the epoch-averaged residuals \addedB{over each frequency band}. \addedA{The two events are clearly seen in the timeseries though most prominently in the 820~MHz band.} The second event dip is constrained by the \addedB{higher-cadence} 1400~MHz data to be between MJD 57508 and 57512. The perturbation amplitudes were $\approx\!-3.0,-1.3$, and $-0.7~\mu\mathrm{s}$ at 820, 1400, and 2300~MHz, respectively; interstellar-propagation effects generally have increased amplitude at lower frequencies \citep[see e.g.,][]{handbook}.

\subsection{Traditional Timing Model Fitting with Per-Epoch DM Variation Estimates}
\label{sec:traditional}

\addedA{In this section, we describe a traditional timing model that includes time-varying dispersive delays along with the standard achromatic timing model (spin-down, binary motion, etc.) and red-noise terms. We now refer to general time delays proportional to $\nu^{-2}$ (traditional DM delays) as $\Delta t_{2,1400}$, where 2 refers to the frequency-dependence index of $-$2 and 1400 refers to a fiducial frequency of 1400~MHz. Thus the observed TOA at frequency $\nu$ is
\be
t_\nu = t_\infty + \Dttwo\left(\frac{\mathrm{1400~MHz}}{\nu}\right)^2 + \epsilon_\nu
\ee
where $t_\infty$ describes the ``infinite-frequency'' arrival time, i.e., the achromatic delay terms, and $\epsilon_\nu$ is the TOA measurement uncertainty. Here we will assume that the $\Dttwo$ delays are attributed entirely to dispersive delays, i.e., the estimated DM (written with a carat to denote it is a proxy for the true DM) is related to the delay by $\Dttwo = K\widehat{\DM}/(1400~\mathrm{MHz})^2$ where the dispersion constant $K \approx 4.149 \times 10^9$~$\mu$s~MHz$^2$~pc$^{-1}$~cm$^3$ in observationally convenient units.
}

We fit a full timing and noise model to TOAs \addedA{\addedB{across} all frequencies} using the \textsc{Tempo}\footnote{\url{http://tempo.sourceforge.net}} timing package \citep{Tempo} with the \textsc{ENTERPRISE}\footnote{\url{https://github.com/nanograv/enterprise}, which uses \textsc{Tempo2} \citep{Tempo2,Tempo2ASCL} via libstempo: \url{https://github.com/vallis/libstempo}} analysis code to estimate the noise parameters. We fit all parameters described \addedA{previously in \S\ref{sec:fixedmodel} and} in NG11. \addedA{The DM model assumes a step-wise value of DM (DMX), i.e., constant over rolling periods of up to six days as in NG11}. We used an F-test described by \citet{NG9yr} to determine if new parameters should be added but none were found to be significant.

Figure~\ref{fig:DMX} shows the $\Dttwo\propto\!\nu^{-2}$ timeseries modeled by \addedA{our step-wise DM model} and the trajectory of the pulsar across the sky, with decreasing values shown by darker colors, similar to the depiction in \citet{NG9DM}.

\subsection{Multi-Component Chromatic Fitting}
\label{sec:enterprise}

\addedA{Now, we describe a timing model that incorporates time-variable chromatic behavior that is more flexible than the model used in \S\ref{sec:traditional}. In addition to time delays proportional to $\nu^{-2}$, we also added a second set of delays proportional to $\nu^{-4}$, which we label $\Delta t_{4,1400}$ using the previous notation.  We continued to include the same achromatic terms as before. Thus, the observed TOA at frequency $\nu$ becomes
\ba
t_\nu & = & t_\infty + \Delta t_{2,1400}\left(\frac{\mathrm{1400~MHz}}{\nu}\right)^2 \nonumber \\
& & +\Delta t_{4,1400}\left(\frac{\mathrm{1400~MHz}}{\nu}\right)^4 + \epsilon_\nu.
\ea
}

\addedA{Our multi-component} model \addedA{of the chromatic variations was} fit \addedA{to all of the TOAs \addedB{directly}} using \textsc{ENTERPRISE}. \addedA{Since we no longer fit for chromatic delays at every epoch, this model contains far fewer parameters than the previously-described model}. We included the usual timing and noise parameters in the \addedA{fit described in \S\ref{sec:traditional}}, along with a \addedA{\addedB{generic}} power-law Gaussian process $\propto\!\nu^{-2}$ \addedB{delays} describing interstellar turbulent variations \addedA{(a Gaussian process refers to a general way to model the timeseries rather than the distribution of density inhomogeneities)}, quadratic polynomial DM \addedB{over the whole timeseries} \addedA{(an actual dispersive-only term, for relative Earth-pulsar motions; \citealt{LamDMt})}, yearly DM sinusoid \addedA{LOS motions from a potential DM gradient in the ISM}, solar wind DM component \citeMadison, and two negative $\nu^{-2}$ step functions with exponential decays back to the original value \addedA{empirically describing the events.} \addedA{We denote the above as Model A. \addedB{Each of the above are meant to describe separate astrophysical phenomena but note that there will be large covariances between the terms; future work is needed in determining best practices for modeling longer-term chromatic variations in TOA timeseries.} In addition to these \addedB{Model A} components, we added either a} power-law Gaussian process $\propto\!\nu^{-4}$ component \addedA{that accounts for possible scattering or refractive variations \citep{fc90} over the entire timeseries (Model B) or a $\propto\!\nu^{-4}$ exponential decay term with the same start time and decay constant as $\nu^{-2}$ delay term for the second event (Model C); our earlier data around the time of the first event are not sensitive to multiple chromatic components because of the small bandwidths observed for the individual bands as previously described (see also \citealt{NG9yr})}. \addedB{The components for all three models are described in Table~\ref{table:models}.}

\begin{deluxetable}{lc|ccc}
\tablecolumns{5}
\tablecaption{Model Parameters Used in Multi-Component Chromatic Fitting\label{table:models}}
\tablehead{
\multicolumn{2}{c}{Model Components} & \multicolumn{3}{c}{Models}\\
\colhead{\vspace{-0.2cm}Description of} & \colhead{Frequency } & \colhead{} & \colhead{} & \colhead{}\\
\colhead{Time Dependence} & \colhead{Dependence} & \colhead{A} & \colhead{B} & \colhead{C}
}
\startdata
Power-law Gaussian process & $\nu^{-2}$ & $\times$ & $\times$ & $\times$ \\
Quadratic polynomial  & $\nu^{-2}$ & $\times$ & $\times$ & $\times$\\
Yearly sinusoid  & $\nu^{-2}$ & $\times$ & $\times$ & $\times$\\
Solar wind & $\nu^{-2}$ & $\times$ & $\times$ & $\times$\\
\vspace{-0.2cm}Negative step function with & & & & \\
exponential decay for first event & $\nu^{-2}$ & $\times$ & $\times$ & $\times$\\
\vspace{-0.2cm}Negative step function with & & & & \\
exponential decay for second event  & $\nu^{-2}$ & $\times$ & $\times$ & $\times$\\
Power-law Gaussian process & $\nu^{-4}$ &  & $\times$ & \\
\vspace{-0.2cm}Negative step function with & & & & \\
exponential decay for second event  & $\nu^{-4}$ & & & $\times$\\
\enddata
\end{deluxetable}

Figure~\ref{fig:enterprise} shows the $\nu^{-2}$ and $\nu^{-4}$ \addedB{delays} \addedA{for \addedB{Model B}}. Our fit exponential components have amplitudes $\Dttwo\approx1.8$ and $1.1~\mathrm{\mu s}$ and decay timescales $\approx62$ and $25$~days \addedB{for the two events}, respectively. \addedA{Again, while our early data are insensitive to multiple chromatic components}, we saw variations in the $\nu^{-4}$ component for MJD$\gtrsim$56000; the $\nu^{-2}$ power-law process was consistent with a Kolmogorov turbulent fluctuation spectra \addedB{i.e., with spectral index $\approx -8/3$ \citep[e.g.,][]{LamDMt}}. Note we parameterized the non-$\nu^{-2}$ delays as a $\nu^{-4}$ component but they need not have this index or even have a power-law dependence \addedA{\citep[see e.g.,][]{Cordes+2017}}. Replacing the $\nu^{-4}$ term with an alternate power law (i.e. $\nu^{-3}$ or $\nu^{-5}$) \addedA{were marginally less-favored for both Models B and C using the differences in the Bayesian Information Criteria \citep[BIC;][]{BIC}. The $\Delta$BIC was $\approx1$ for both $\nu^{-4}$ models over Model A. Such a value describes weak evidence for $\nu^{-4}$ delays even though they are preferred marginally. Again, any events in our data need not take on the phenomenological forms that we have chosen in Models B and C;} future analyses should utilize more robust model selection for proper astrophysical inferences. Higher-order frequency terms can indicate LOS refraction or scattering \citep{fc90,LamDMt}, which we discuss in \S\ref{sec:interpretations}.

\section{Pulse-Profile Analyses}
\label{sec:otherradio}

\addedA{In addition to timing analyses, we performed analyses directly on the pulse profiles. We looked for changes both in interstellar scattering parameters and intrinsic pulse shape variability with time.}

\subsection{Changes in Flux and Scintillation Parameters}

We generated dynamic spectra using the GUPPI/PUPPI \addedA{large-bandwidth} data and calculating the flux density for \addedA{pulses in} each time-frequency bin using \textsc{PyPulse}\footnote{https://github.com/mtlam/PyPulse} \citep{PyPulse}. Since we only had GASP/ASP data covering the first event, we did not have sufficient bandwidth for which to estimate diffractive scintillation parameters. We used a template-matching approach similar to that used to generate TOAs but \addedA{for determining} the pulse amplitudes \addedA{used to generate the dynamic spectra}. We generated the TOAs described in \S\ref{sec:observations} using data summed over an entire 20-30~min observation. For this scintillation analysis, we summed over 1-2 minutes for increased time resolution \citep{NG9WN}.

We used the 2D autocorrelation functions (ACFs) of dynamic spectra to estimate scintillation parameters. The scintillation timescale (at our observing frequencies) is of \addedA{the} order \addedB{of} our observation lengths and thus we could not measure it. For our 820- and 1400-MHz observations, we estimated the scintillation bandwidth $\Dnud$ \addedA{using the half width at half maxima of the ACFs along the frequency lag axis \citep{Cordes1986}. We estimated the scattering timescale via the relationship $\taud=1.16/(2\pi\Dnud)$ as well} \citep{cr98}.

Following \citet{Levin+2016}, we ``stretched'' the dynamic spectra to 1400~MHz to remove the effect of frequency-dependent scintle size evolution across the band. To build S/N, we averaged the ACFs in 100-day bins, corresponding to the approximate length of the events, and then estimated $\Dnud(t)$ and $\taud(t)$. At our current sensitivity, we saw no significant variations in $\taud$ over the time of the second event. 

We also computed the average flux density over each observation to look for longer-timescale refractive interstellar scintillation variations; the estimated refractive timescale is $\approx$3.5~days \citep{sc1990,Keith+2013,Levin+2016}. The variations were consistent with diffractive interstellar scintillation only and we did not have sufficient \addedA{time resolution/cadence} to separate the refractive and diffractive components.

We did not have sufficient observation lengths (resolution in conjugate time) to see scattering material in the form of scintillation arcs in secondary spectra (2D Fourier transform of the dynamic spectra) in the manner of \citet{Stinebring+2001}. We re-analyzed an eight-hour GBT observation from a 24-hour campaign targeting PSR~J1713+0747 \citep{Global1713}. The increased observation length provided finer resolution but we did not see clear scintillation arcs though there is some notable power off the axis where conjugate time is zero. We also measured $\taud$ from the ACF, which was consistent with the standard NANOGrav observations.

\subsection{Pulse-Shape Variability}

Long-term temporal pulse-profile variations have been seen in many canonical pulsars \citep{Lyne+2010,Palfreyman+2016} but only one recycled millisecond pulsar, PSR J1643$-$1224 \citep{Shannon+2016}; these variations will affect the measured TOAs. The timing residuals \addedA{for PSR~J1643$-$1224} showed a similar exponential shape with recovery as the events reported here, although with inverted frequency dependence. We tested whether the observed timing effects are due to changes in the pulse shape. Using the method in \citeBrook for NG11 \addedA{\citep[adapted from][]{Brook+2016}}, we used a Gaussian process to model variations of each 820- and 1400-MHz \addedA{band-averaged} pulse profile across epochs on a per-phase-bin basis. We did not see significant profile modulation per epoch above what is normal from intrinsic frequency-dependent pulse-shape evolution \addedA{modulated by scintillation (which weights the pulse profiles as a function of frequency)} at the event times.


\section{Infrared and Optical Observations}
\label{sec:multiwavelength}

Since we believe the \addedA{possible} non-$\nu^{-2}$ delays arise from phenomena in the ISM, we searched for possible ISM structures that might be associated with the delays.

We examined 2MASS images of the field in the \textit{J/H/K} (1.2/1.6/2.2~$\mathrm{\mu m}$) bands. We also inspected a Palomar \textit{g}-band image and IRIS (Improved Reprocessing of the IRAS Survey) 12/25/100~$\mathrm{\mu m}$ images. We did not detect any interstellar structures that could be responsible for pulse dispersion changes along the LOS. No known H{\textsc{II}} region along the LOS has been seen previously \citep{Anderson+2014}. We examined Catalina Sky Survey lightcurves \citep{Catalina} but the closest source is separated by $\sim\!36''$, with no statistically significant brightness variations.

\begin{figure}[t!]
\hspace{-7ex}
\includegraphics[width=0.58\textwidth]{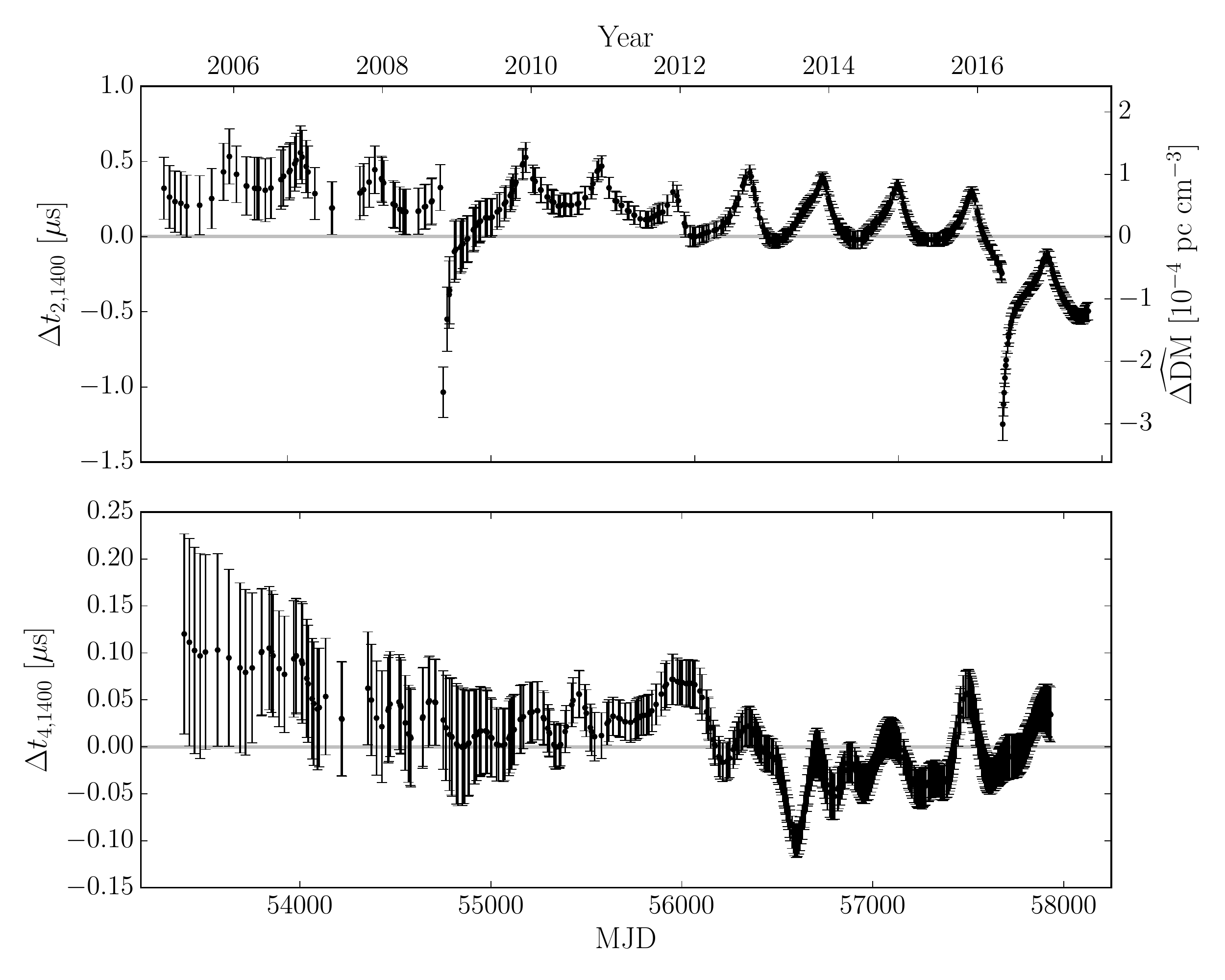}
  \caption{\footnotesize \addedA{Estimated multi-component chromatic model} timeseries of $\Dttwo\!\propto\!\nu^{-2}$ delay (top) and $\Dtfour\!\propto\!\nu^{-4}$ delay (bottom) scaled to 1400~MHz. See Figure~\ref{fig:DMX} regarding $\Delta\widehat{\DM}$. Errors show the 68.3\% confidence intervals of the \addedA{model} realizations after mean subtraction. \addedA{Note the tick marks denote time in MJD and years for the bottom and top ticks of each panel, respectively.}}
\label{fig:enterprise}
\end{figure}

\citet{br2014} measured an upper limit on the $\Halpha$ flux of $5.9\times10^{-5}~\mathrm{cm^{-2}~s^{-1}}$ within $\sim\!0.9''$ of the pulsar position. With assumptions about the Galactic warm neutral and ionized media structure, and pulsar velocities, they estimated the expected flux if pulsars are producing bow shocks; for PSR~J1713+0747 they had sufficient sensitivity to detect this flux. We are unaware of additional $\Halpha$ surveys of sufficient angular resolution to probe LOS structures.


\section{Interpretations}
\label{sec:interpretations}

We discuss the possibility of the events being independent and unassociated or linked by some structure near the pulsar as possible interpretations for the observed timing perturbations. \addedA{We also} consider a broad plasma lens interpretation.

\subsection{Independent Events}
\label{sec:independent}

We considered whether the events were due to two arbitrary independent processes causing purely-dispersive DM variations at the level of $6\times10^{-4}~\mathrm{pc~cm^{-3}}$ with timescales of $\sim$100 days; we saw no other such events in NG11. Using each NG11 DMX timeseries, we summed the total time over the PTA where the median DM errors were smaller than $1.2\times10^{-4}~\mathrm{pc~cm^{-3}}$, corresponding to an event $\SN\gtrsim5$. For 27 pulsars and 187 pulsar-years of total time, including the 12.5 years for PSR~J1713+0747, we found a Poisson rate of $0.13-3.9\times10^{-2}~\mathrm{yr^{-1}}$ at the 95\% confidence level, or $0.02-0.48$ events expected in a given 12.5~year timespan following \citet{Gehrels1986}. We concluded that two independent events would not have been observed in the single timeseries alone given the event rate.

\subsection{Local Structure}

We next \addedA{considered the possibility} that a single structure local to the pulsar crosses the LOS since the events are unlikely to be caused by independent structures as previously stated. The bottom panel of Figure~\ref{fig:DMX} shows the path of the pulsar and the event start times; it is unlikely that the LOS passes through \addedA{an assumed} interstellar structure very close to the pulsar unless the structure is orbiting the pulsar \addedB{since the pulsar's apparent position has moved across the sky}. In addition, a cloud causing a purely dispersive delay would only increase the observed DM, not decrease it.

To check for periodicities of \addedA{such structures}, we examined the \DDM~timeseries of \citet{Zhu+2015} for PSR~J1713+0747, which extends back to 1998. Given the 7.6-year interval between the events we observed, if the events were periodic, the previous event would have been at 2011.2. However, no such event is evident in  the timeseries of \citet{Zhu+2015}. We note a low-significance DM dip event in that data set in early 2002, but it is far from the predicted time.

A gap or void through the pulsar's local medium can also decrease the DM. The \DDM~timeseries can show dips as seen in \S7.2 of \citet[see Figure 9 of that paper]{LamDMt} if the pulsar moves through both local high- and low-density structures oriented in specific ways along the LOS. Looking at the timescale $t$ for the events to recover by $\DDM\approx6\times10^{-4}~\mathrm{pc~cm^{-3}}$ to their initial values after approximately 100-200 days, the local $n_e\sim\DDM/(v_pt)\sim10-20~\mathrm{cm}^{-3}$, a large value for the interstellar medium \citep[ISM; ][]{Draine} but one that is marginally consistent with the value estimated for PSR~J1909$-$3744 and possibly PSR~J1738+0333 \citep{LamDMt,NG9DM}. Since the pulsar's 3D velocity $\vec{v}_p$ is not known, we \addedA{first} assumed above that the pulsar moves purely radially towards the Earth (to produce a negative \DDM) with a fiducial velocity $v_{p,\parallel}=100$~km/s and ignored the contribution of the Sun's motion through its own local environment; \addedA{for reference} note that the pulsar's transverse velocity is $v_{p,\perp} = 36.4$~km/s, estimated from the parallax and proper motion.

If the pulsar is moving towards us such that high-density structures explain the event recoveries, then the pulsar must have a transverse velocity component through low-density structures to explain the rapid dips \citep[see \S7.2 of][]{LamDMt}.  The short timescale for the onset of the second event ($\lesssim$4~days; see \S\ref{sec:observations}) implies that $n_e$ must drop drastically within a distance of $\lesssim$0.08~AU as the pulsar moves through the material. Since the transverse motion causes a second-order pulsar-Earth distance change compared with the radial motion \citep{LamDMt}, \DDM~is related to the change in $n_e$ as 
\be
\DDM\approx\Dne\left(\frac{L}{D_p}\right)\frac{\left(v_{p,\perp}t\right)^2}{2D_p},
\ee
with pulsar distance $D_p$, depth of the material at the pulsar $L$ ($L/D_p$ is the LOS filling factor), pulsar transverse velocity $v_{p,\perp}$, and again time $t$. Rearranging and substituting in values, we have
\be
\Dne\approx-9\times10^{12}~\mathrm{cm^{-3}}\left(\frac{L}{D_p}\right)^{-1}.
\ee
Since $L/D_p$ must be less than one, the $n_e$ change is not physically plausible, i.e., there is no possible low-density region compared to the typical ISM that can account for the sudden dips we see in the DM timeseries. Therefore, we rule out the possibility of a non-periodic local structure near the pulsar as the cause of the events.

\subsection{Plasma Lensing}

\addedA{Lensing of the pulsar emission by compact interstellar electron-density-variation regions (``plasma lensing'')} appears to be compatible with our observations and might explain the \addedA{possible} non-$\nu^{-2}$ chromatic timing variations.  Here we will \addedA{discuss the mechanism and} explore the implications of \addedA{the possible connection with our observations}. 

\added{Plasma lenses have been considered previously in the form of a Gaussian cloud \citep{Clegg+1998,Cordes+2017} or folded current sheets \citep{sp2017} and are consistent with observed ``extreme scattering'' events \citep{Fiedler+1987,Coles+2015,Bannister+2016,Kerr+2017}. These over- or under-densities of interstellar free electrons can alter the TOAs in a frequency-dependent manner and modulate pulse fluxes.}

Three phenomena can impose chromatic time delays on the detected pulsar signals: the dispersive delay from propagation through free electrons, the geometric delay from refraction increasing the path length, and the barycentric delay due to correction for the angle-of-arrival variation \citep{fc90,LamDMt}. Discrete structures able to cause these delays will likely produce multiple images at some epochs; the mapping from time to transverse physical coordinates in the plane of the lens is nonlinear and involves the lens equation \citep{Cordes+2017}. Since the two events are asymmetric in time, the lens itself may have an asymmetric structure. There are likely many non-unique solutions to the lens structures; therefore, we leave the analysis to future work. The lensing structure may be embedded in larger-scale material that may have affected a broader range of our TOAs and may in the future cause other arrival-time perturbations. We note that the TOA advances of the events with respect to the surrounding epochs suggest that the lens is related to some under-density.

For a lens with dispersion measure $\DM_l\sim n_eL$ and size $L$ at a distance $D_l$ from the observer, the ratio of the geometric delay to the dispersive delay is \citep{fc90,LamDMt}
\ba\vspace{-3ex}
\frac{t_{\mathrm{geo}}}{t_{\DM}}&\sim&\left[\frac{D_l\left(1-D_l/D_p\right)\lambda^4r_e^2\left(\DM^\prime_l\right)^2}{8\pi^2 c}\right]/\left[\frac{\lambda^2r_e\DM_l}{2\pi c}\right]\nonumber\\
&\sim&\frac{D_l\left(1-D_l/D_p\right)\lambda^2r_en_e}{4\pi L\zeta^2}\nonumber\\
&\approx&4.2\times10^{-3}\zeta^{-2}\left(\frac{D_l}{\mathrm{kpc}}\right)\left(1-\frac{D_l}{D_p}\right)\times\nonumber\\
&&\left(\frac{n_e}{\mathrm{cm^{-3}}}\right)\left(\frac{\nu}{\mathrm{GHz}}\right)^{-2} \left(\frac{L}{\mathrm{AU}}\right)^{-1},
\label{eq:tgeo_vs_tDM}
\ea
with DM spatial gradient $\DM^\prime_l\sim n_eL/(\zeta L)\sim n_e/\zeta$, electromagnetic wavelength $\lambda$, classical electron radius $r_e$, and depth-to-length aspect ratio of the lens $\zeta$. The barycentric to dispersive delay ratio is
\ba
\frac{t_{\mathrm{bary}}}{t_{\DM}}&\sim&\left[\frac{\left(1-D_l/D_p\right)\lambda^2\Rearth r_e\DM_l^\prime}{2\pi c}\right]/\left[\frac{\lambda^2r_e\DM_l}{2\pi c}\right]\nonumber\\
&\sim&\frac{\left(1-D_l/D_p\right)\Rearth}{L}\nonumber\\
&\approx&\left(1-\frac{D_l}{D_p}\right)\left(\frac{L}{\mathrm{AU}}\right)^{-1},
\label{eq:tbary_vs_tDM}
\ea
where $\Rearth$ is the Earth-Sun distance. Note that the barycentric delay can be negative depending on the DM spatial gradient and orbital position of the Earth; the 7.6-year time between events means that the Earth was on opposite sides of its orbit.

If we assume the events are due to caustic crossings at each end of a lens, then $L$ must be $\lesssim v_{\mathrm{eff}}t_{\rm cross}$, with crossing time $t_{\rm cross}$ and effective velocity \citep{cr98,Cordes+2017}
\be
\vec{v}_{\mathrm{eff}}=\left(1-\frac{D_l}{D_p}\right)\vec{v}_{p,\perp}+\left(\frac{D_l}{D_p}\right)\vec{v}_{e,\perp}-\vec{v}_{l,\perp},
\ee
\addedB{where the} pulsar, Earth, and lens velocities transverse to the LOS are $\vec{v}_{p,\perp}$, $\vec{v}_{e,\perp}$, and $\vec{v}_{l,\perp}$, respectively. For a stationary lens situated halfway between the Earth and pulsar, with the pulsar and Earth velocities aligned, $v_{\mathrm{eff}}\approx 18~\mathrm{km/s}$ given the proper motion, implying $L\lesssim 18~\mathrm{km/s}\times 7.6~\mathrm{yr}\approx 30~\mathrm{AU}$.

Refraction from such a lens may explain the \addedA{possible non-$\nu^{-2}$} delays \citep[Eq.~\ref{eq:tgeo_vs_tDM} or see][]{fc90} from our multi-component fitting \addedB{but again} we do not see significant changes in the scattering parameters \addedB{from the small $\Delta$BIC value though we have only tested a small number of possible models and lensing can produce non-$\nu^{-2}$ delays that have complex dependencies in frequency and time \citep{Cordes+2017}; assuming that Model B is correct then the fluctuations in Figure~\ref{fig:enterprise} do show significant temporal variations}. If \addedB{refraction does explain the observed delays}, then our crude analysis suggests that the change in $n_e$ with respect to the surrounding medium is large and/or the lens is highly compact for the geometric delay to become important \addedA{as per Eq.~\ref{eq:tgeo_vs_tDM}}; however the barycentric delay becomes important for compact lenses as suggested above in Eq.~\ref{eq:tbary_vs_tDM}. A more detailed analysis is outside the scope of this paper. Any such structures are diffuse enough however to be undetected in the multiwavelength observations discussed in \S\ref{sec:multiwavelength}.

\section{Discussion}
\label{sec:discussion}

\addedA{In our analysis,} we \addedA{describe} the observed timing events as \addedA{possibly being} due to a single \addedA{lensing} structure in the ISM. We showed that these events likely cannot be interpreted as being caused by the dispersive delay alone. In general, \addedA{estimates of a $\nu^{-2}$ delay from TOAs should only be considered a proxy for a truly dispersive delay}.

It is notable that one of the pulsars most sensitive to timing perturbations shows evidence of a plasma lens with such large arrival-time amplitudes. As PTAs observe more pulsars over longer times, searches for similar chromatic-delay events may allow us to find a larger population of plasma lenses. Sensitivity in the scintillation parameters can be yielded by cyclic spectroscopy and may help probe future chromatic-timing events \citep{Demorest2011,Stinebring2013}. Quasi-real-time processing of TOAs can allow for faster identification of such features in our data and provide us with the ability to adjust observations accordingly to provide better temporal and frequency coverage of future similar events.

\acknowledgments

\addedB{
{\it Author contributions.} M.T.L. performed the primary analyses and prepared the majority of the text and figures. J.A.E., G.G., M.L.J., J.S.H., P.R.B., J.E.T., and S.C. assisted in different analyses associated with this work. J.M.C, V.G., T.J.W.L, D.R.M., M.A.M., and D.R.S contributed useful discussions regarding the development of the framework. Z.A., H.B., P.R.B., H.T.C., M.E.D., P.B.D., T.D., J.A.E., R.D.F, E.C.F, E.F., N.G., P.A.G., M.L.J, M.T.L, D.R.L, R.S.L, M.A.M., C.N., D.J.N, T.T.P., S.M.R., R.S., I.H.S, K.S, J.K.S, S.J.V., and W.W.Z. all ran observations and/or otherwise assisted in the creation of the preliminary NANOGrav 12.5-year data set and pipeline. M.T.L. and M.E.D. worked more specifically on the generation of the initial timing model for PSR~J1713+0747 used in this work.
}

{\it Acknowledgments.} \addedA{We thank Daniel Evans for comments on the infrared/optical observations.} The NANOGrav Project receives support from NSF Physics Frontiers Center award number 1430284. Pulsar research at UBC is supported by an NSERC Discovery Grant and by the Canadian Institute for Advanced Research. The Arecibo Observatory is operated by SRI International under a cooperative agreement with the NSF (AST-1100968), and in alliance with Ana G. M\'{e}ndez-Universidad Metropolitana, and the Universities Space Research Association. The Green Bank Observatory is a facility of the National Science Foundation operated under cooperative agreement by Associated Universities, Inc. Part of this research was carried out at the Jet Propulsion Laboratory, California Institute of Technology, under a contract with the National Aeronautics and Space Administration. WWZ is supported by the CAS Pioneer Hundred Talents Program and the Strategic Priority Research Program of the Chinese Academy of Sciences, Grant No. XDB23000000.

\software{PSRCHIVE \citep{Hotan+2004,vSdo2012}, Tempo \citep{Tempo}, Tempo2 \citep{Tempo2,Tempo2ASCL}, libstempo (\url{https://github.com/vallis/libstempo}), ENTERPRISE (\url{https://github.com/nanograv/enterprise})}

\end{document}